**The Evolving Block Universe and the Meshing Together of Times**

**George F R Ellis, University of Cape Town**[1]

**Abstract:** *It is proposed that spacetime should be regarded as an evolving block universe, bounded to the future by the present time, which continually extends to the future. This future boundary is defined at each time by measuring proper time along Ricci eigenlines from the start of the universe. A key point is that physical reality can be represented at many different scales: hence the passage of times may be seen as different at different scales, with quantum gravity determining the evolution of space time itself but quantum field theory determining the evolution of events within spacetime .The fundamental issue then arises as to how the effective times at different scales mesh together, leading to the concept so global and local times.*

**1: Time as an Illusion: The Block Universe**

This paper is one of a series of papers[1-4] developing a proposal for an Evolving Block Universe picture of spacetime, instead of the usual proposal of a Block Universe (BU) [5-10], summarized in Section 1.2. What is novel in this paper is a new set of answers to criticisms of the EBU proposal, and development of the way the multi-scale structure of physical reality relates to it.

**1.1. The basic criticisms of the EBU**

The two key criticisms I have received from an anonymous referee are the following:

- "*The author gives the usual motivations for rejecting the block universe (BU) implications of relativity such as contradicts experience and other areas of science, e.g., thermodynamics. As the author is well aware countless defenders of BU such as Huw Price have pointed out that BU is perfectly consistent with experience and entropy.*"

I shall point out below (Section 1.3) that this claim of consistency can only be sustained by ignoring much evidence at both the macro and micro scales. In particular, this view does not take quantum physics seriously.

- "*Relativity (both SR and GR) plus a few innocuous assumptions, strongly implies a BU. There are only two ways out if one doesn't like BU: 1) add something new to relativity such as a preferred frame or 2) reject realism about Minkowski spacetime.*"

Firstly, my program specifically introduces a preferred frame (Section 2.2), and so satisfies 1). Secondly, as regards 2), Einstein's General Theory of Relativity is the valid classical theory of gravity and spacetime (section 2.3). The foundation of that well established theory is that Minkowski spacetime does *not* describe the real universe; it exists as a Platonic mathematical entity, but not in physical reality. Thus I reject physical realism about Minkowski spacetime, which Einstein showed to be wrong.

---

[1] Email: George.ellis@uct.ac.za

## 1.2 The Block Universe

There is a prevalent view in fundamental physics that space and time are best described as a 4-dimensional spacetime which represents all the places and all the times that ever exist as a single unchanging entity. There is no essential difference between the past and the future, because there is no present time defined to separate them; they cannot be distinguished from each other, so there is no meaningful present. Without an objective present, time does not flow in any real sense: the passage of time is an illusion.[5-10]

The underlying dynamical idea is that given data at an arbitrary time, everything occurring at any later or earlier time can be uniquely determined from that initial data by time reversible Hamiltonian dynamics, which is assumed to be the basis of dynamics of physics in general and of gravitation in particular. The future and the past are both uniquely predictable from the present because one can predict the state $S(\mathbf{x},t)$ equally to the past and the future from data given at an initial time $t_0$:

$$H: S(x,t_0) \rightarrow S(x,t_1) \quad \text{for all } t_1. \tag{1}$$

But there is nothing special about $t_0$: it is an arbitrary choice. Consequently, nothing can be special about any particular moment; there is no special "now" which can be called the present.

**The Block Universe:** Such a view can be formalized in the idea of a Block Universe, where space and time are represented as merged into an unchanging spacetime entity, with no particular space sections identified as the present and no evolution of spacetime taking place. The universe just is: a fixed 4-dimensional spacetime block, representing all events that have happened and that ever will happen.[9,10] Past, present and future are equal to each other, for there is no surface which can uniquely be called the present. This implicitly embodies the idea that time is an illusion: time does not "roll on".

## 1.3 The problems

We represent physical reality in terms of different scales of description, with larger scales emergent from smaller scales through coarse graining[4,11], but also with higher levels influencing lower level events in many ways [4,11-14]. One can consider a great many levels of description, but for present purposes a 5-scale description will suffice, as shown in Table 1.

| Scale 4 | Cosmology |
|---------|-----------|
| Scale 3 | Astrophysical structures |
| Scale 2 | Macro level: daily life |
| Scale 1 | Micro level: quantum physics |
| Scale 0 | Quantum gravity |

**Table 1: levels of Description**

The issue then is on which scales the BU description might be accurate, and how the descriptions at different scales relate to each other. A key point is that spacetime curvature is determined at Scale 3, because smaller scale entities have a negligible effect on this curvature. Events within spacetime are determined by the interaction between entities at scales 1 and 2.

**Scale 2: Macro problems**

At the macro level, dynamics is almost always not time-reversible: Hamiltonian development is in flagrant contradiction to our experiences in everyday life, as well as all of biology and biochemistry. This is the profound content of the Second Law of Thermodynamics, a fundamental feature of the macro world of physics, chemistry, and biology.[15-17].

Wine glasses fall and break, books get burnt and their ashes get scattered to the wind, species die out and traces of most of them are lost. Furthermore chaotic dynamics amplifies micro fluctuations to the macro scale. At the macro scale the reversible dynamics (1) is mostly not true: conditions in the past do not uniquely determine the future, or vice versa. Examples abound: the classic example is weather forecasting, other examples are where and when lightning will strike, when automobile accidents will happen, when shares on the stock market will rise, etc. One only knows what will happen when it happens, it is not predictable. Hamiltonian dynamics does not apply. Hence a block universe representation is not appropriate at that scale.

However the idea then proposed[17] is that nevertheless there is indeed reversible Hamiltonian dynamics at the micro level, but through coarse graining it leads to irreversible physics at the macro level, because micro information is lost through the coarse graining. The Second Law of Thermodynamics arises through special initial conditions, but that the dynamics is actually time reversible, it just that we don't see this fundamental feature because our senses are too coarse.

If we could accurately reverse all the micro velocities, the macro level description would run in reverse: the fragments of glass on the floor would reassemble into the intact wine glass, the ashes of the burnt book would come back on the wind and recreate the pristine object. The true picture is time reversible, it's just our coarse macro vison that prevents us seeing this. Even if the BU is not a good description at the macro level, it is indeed good at the micro level, so a BU picture is justified at that level and represents the true situation.

This view is however not applicable to cases such as deleting memory files in a digital computer. Once they are overwritten, records are irretrievably lost; no trace is left, information is gone. This is based in the physics of read and write operations in a digital computer, see pages 431-460 in[18], based in the properties of registers (pages 354-364) because of latch property and the underlying basis in flip flops (page 371). And in the end this irrevocable irreversibility, typical of all adaptive selection processes, is based in the properties of the underlying quantum physics.

**Scale 1: Micro problems**

It is often stated that because quantum physics is based in unitary evolution of the wave function, fundamental physics is Hamiltonian and information is never lost. The following quote by Fabbri and Salas[19], made in the context of the debate on the black hole information paradox, is typical:

> *The type of radiation emitted does not allow the recovery of the information about the star from which the black hole was created. Therefore, with the disappearance of the black hole this information will be lost forever. But this is forbidden by the basic principles of quantum mechanics itself."*

**No collapse** The authors therefore believe in a version of Quantum Mechanics where the non-unitary irreversible process of collapse of the wave function [20-24], where information is indeed lost (see Section 3 below), never occurs. Thus in their version measurements never take place. But then the quantum probabilities that the wave function is supposed to represent are never realized in physical reality if no collapse of the wave function ever takes place.

Please consider Figure 1, showing the classic double-slit experiment build-up of an interference pattern of single electrons. Each dot there represents an individual particle arrival, and hence represents non-unitary behavior because in each case a specific classical outcome has occurred (see Section 3). Quantum theory very successfully predicts the statistics of the outcome, the interference patterns that builds up, but it cannot predict where each individual particle will arrive. These individual outcomes are not predictable, as they are essentially uncaused; they are only determined as they happen.

The claim that Quantum mechanics is unitary is conclusively proved wrong by this experiment. There are however three counters to this argument, which I now consider in turn.

**Ensemble** The first response is that quantum physics is not about individual events, it is about statistics of events. While the quantum calculation cannot predict where each individual particle will arrive, it very successfully predicts the statistics of the outcome and associated energies, scattering angles, and similar measurable properties, without considering the process of wave function collapse. This is all physics needs, so it does not matter that this non-unitary behavior related to individual events occurs. We just need to consider the statistics of the ensemble.

The reply is that there is no ensemble to consider unless the individual events that make up the ensemble occur. Hence no statistics of outcomes exists without the non-unitary behavior represented by the successive appearance of the individual dots on the screen, each of which represents an event where collapse of the wave-function occurs. And it is not always true that only the statistics count: specific collapse events can sometimes have crucial macro outcomes (see the Scale 3 discussion below).

**Decoherence** The second response is that environmental decoherence (see [22]), which diagonalizes the density matrix, gives effective collapse of the wave function.

The reply is that this is not in fact the case: while decoherence indeed diagonalises the density matrix and hence gets rid of entanglement, it does not set all the diagonal elements of the density matrix to zero except one, and so it does not get rid of superpositions. Thus it does not lead to classical outcomes, and so cannot account for the individual events seen in Figure 1.

**Many worlds** The third is the Everett many worlds view (see [22]), where unitarity is preserved by denying the occurrence of wavefunction collapse. Reality is viewed as a many-branched tree, where every possible quantum outcome occurs as superpositions are generated by the unitary dynamics. All possible alternative histories occur, each representing an actual universe.

This proposal raises major philosophical issues relating to testability and to Occam's razor, and there are significant problems to do with the issue of measures, a preferred basis, and how this approach gives the Born probability rule. However for present purposes the key point is that the many worlds proposal does *not* give the block universe – it gives an ever branching spacetime with billions and billions of ongoing bifurcations (see Tegmark[25] Figs.8.10 and 12.2 for the case of just one such bifurcation). This looks nothing like the standard block universe, and the resulting spacetime has never been described in coordinate terms (new coordinates would need to be added every time a bifurcation take place).

This proposal does not support the block universe view, it gives a completely different spacetime picture, which we cannot adequately represent by any spacetime diagram at all. And if we could, there would also be an Evolving Everett Universe version, which would be required to adequately represent the passage of time. I will not consider this proposal further here.

**Scale 3: The Micro-macro connection**

We have seen that real world evolution of events is not unitary on scales 1 and 2. What about scale 3? It is also not unitary on this scale either, because of the current standard model of cosmology, which includes an epoch of inflation—an extraordinarily rapid accelerating expansion for a very brief period - at very early times.[26-28]

The situation is represented in Figure 2. The key point is that quantum fluctuations in the inflationary epoch generated inhomogeneities on the surface of last scattering of matter and radiation, which then provided the seeds for growth of inhomogeneities by gravitational collapse. This was the origin of large-scale structures such as clusters of galaxies at the present time. But because they were quantum fluctuations, complete knowledge of the state of the universe at the start of inflation does not determine what specific fluctuations occurred on the surface of last scattering, and hence what large scale structures will exist in the universe today. The specific outcomes of these Gaussian random processes were only determined as they occurred. The evolution of matter – and hence of spacetime geometry – at Scale 3 has been non-unitary over cosmic timescales because of this extraordinary micro-macro connection caused by the mechanism of inflation. In effect, the uncertainty of individual events shown in Figure 1 occurs writ large in the sky because of the cosmological context depicted in Figure 2.

**Conclusion**: The BU spacetime model is inadequate on scales 1, 2, and 3. We must be able to do better, adequately representing the fact that that in the real universe, unpredictable things happen at each of these scales as time progresses. In actual fact, physical outcomes do not proceed in a unitary way.

2. **The evolving block universe**

By contrast to the usual Block Universe view, one can suggest that the true nature of spacetime is best represented as an Evolving Block Universe (EBU), a spacetime which grows and incorporates ever more events, "concretizing" as time evolves along each world line.[1-4,29] This is the same as the usual 4-dimensional block universe except that the future boundary of spacetime no longer represents the infinite future: it represents the present time, which is that instant along our world line where at this moment the indefiniteness of the future changes to the definiteness of the past. It continually moves to the future, incorporating ever more spacetime events as time passes.

**2.1 The basic idea** Consider a massive object with two computer controlled rocket engines that move it right or left. Let the computer determine the outcome on the basis of measurements of decay products of radioactive atoms. Then the outcome is unpredictable in principle, because of the foundational quantum uncertainty of the photon emission process.[20-24] If the object is massive enough, it curves spacetime, and so the future spacetime structure is not determinable or predictable from current data. Selection of the specific path taken, and hence the spacetime structure that results, occurs on an ongoing basis as radioactive decays take place in an unpredictable way. The change from uncertainty to certainty takes place at the ever changing present, where the indefinite future becomes the determined past (Figure 3).

The future does not exist in the same sense as the past or the present:
- The past has been determined and is fixed,
- The future is uncertain and still has to be fixed; because of quantum uncertainty, it is not true that the future is determined at the present time;
- The present is where the change takes place. It is crucially different from the past and future, and indeed separates them: it is the future boundary of the determinate spacetime region.

Thus the EBU (Figure 4) is exactly the same as the block universe, except it has a future boundary, namely the present, which is not static: it continually extends in the future direction of time. Spacetime itself is growing as time passes.[1-4,29] This obviously represents the passing of time in a more satisfactory way than the usual block universe.

**2.2 The problem of the present time** The primary problem with this proposal is the claimed unique status of "the present" in the EBU - the surface where the indeterminate future is changed to the definite past at any instant. It is a fundamental feature of Special Relativity that simultaneity is not uniquely defined, it depends on the state of motion of the observer.[16,23,30] Hence it is claimed that no preferred present time can exist, hence the block universe model is the only way a spacetime model can take account of this lack of well defined surfaces of instantaneity.

**General relativity** However it is general relativity that describes the structure of space time, not special relativity. Gravity governs space-time curvature, the metric tensor is determined by the matter present through the Einstein field gravitational equations.[31] Because there is no perfect vacuum anywhere in the real universe, inter alia because cosmic blackbody background radiation[26-28] permeates the Solar System and all of interstellar and intergalactic space, space time is nowhere flat or even of constant curvature; therefore there are preferred timelike lines everywhere in any realistic spacetime model. The special relativity argument does not apply: Minkowski Spacetime does not exist in physical reality.

**The irrelevance of simultaneity** Furthermore, simultaneity as usually defined, determined by radar,[30] is irrelevant to physical causation. Consider the Mars Rover, controlled from Earth. There is a communication time delay between Earth and Mars that is about 20 minutes on average. What matters physically is E1 (emission of a control signal from Earth), E2 (reception at Mars and emission of reply), and E4 (reception of this reply back at Earth). Which event S is simultaneous with E2 has no physical significance: it only has psychological value.

**Resolution:** Physically, things happen along timelike worldlines rather than on spacelike surfaces. What matters physically is proper time measured along preferred timelines $x^i(v)$ by perfect clocks, determined in terms of the metric tensor $g_{ij}(x^k)$ by the basic formula[30,31]

$$\tau = \int (-ds^2)^{1/2} = \int (-g_{ij}(dx^i/dv)(dx^j/dv))^{1/2}\, dv. \qquad (2)$$

**Time of determination:** Start at the beginning of time, measure proper time $\tau$ given by (2) along fundamental world lines, thereby determining the present instant at time $\tau$ as time passes on each of these preferred fundamental world lines. This happens locally everywhere, determining the present time $\tau$ along each such world line, for each value of $\tau$.

**Resultant surfaces of change: "The Present"** Natural surfaces of constant time are given by this integral since the start of the universe Thus we can propose that[3]

> **The present:** *The ever-changing surface $S(\tau)$ separating the future and past - the 'present' – at time $\tau$ is the surface $\{\tau = constant\}$ given by integral (2) along a family of fundamental world lines, starting at the beginning of space time.*

But is this well-defined, given that there are no preferred world-lines in the flat spacetime of special relativity? Yes indeed, again because it is general relativity that describes spacetime.

**The Preferred Worldlines:** A unique geometrically determined choice for fundamental worldlines is the set of timelike eigenlines $x^a(v)$ of the Ricci tensor on a suitable averaging scale, representing the local average motion of matter[27,32] (they will exist and be unique for all realistic matter, because of the non-zero Cosmic Background Radation and the energy conditions[31] such matter obeys). Their 4-velocities $u^a$(v) = $dx^a(v)/dv$ satisfy

$$\mathbf{T_{ab}\ u^b = \lambda_1\ u^a \Leftrightarrow R_{ab}\ u^b = \lambda_2\ u^a} \qquad (3)$$

where the equivalence follows from the Einstein field equations. Thus we can further propose that[3]

> **Fundamental world lines:** *the proper time integral (2) used to define the present is taken along the world lines with 4-velocity $u^a(v)$ satisfying (3).*

It is well defined in any realistic cosmological model, and will give the usual surfaces of constant time in the standard Friedmann-Lemaître-Robertson-Walker (FLRW) cosmologies.

**What about simultaneity?** In general these surfaces are not related to simultaneity as determined by radar; indeed this is even so in the FLRW spacetimes (where the surfaces of homogeneity are generically not simultaneous according to the radar definition[33]).

The flow lines are not necessarily orthogonal to the surfaces of constant time. This does not matter: no physical phenomena are directly determined by simultaneity in the usual sense. More than that, the surfaces determined in this way are not even necessarily spacelike, in an inhomogeneous spacetime. In that case the implied initial value problem will locally be timelike, and the way it works will need to be rethought.

## 2.3 The Evolution of Space Time

**The metric evolution**: So if the metric tensor determines proper time, what determines the metric tensor? The Einstein field equations, of course! [31]

Following the ADM formulation[34], the first fundamental form (the metric) is represented as

$$\mathbf{ds^2 = (-N^2 + N_iN^i)dt^2 + N_i dx^j dt + g_{ij} dx^i dx^j} \qquad (4)$$

where i, j = 1, 2, 3. The lapse function N(x) and shift vector $N^i(x)$ represent coordinate choices, and can be chosen arbitrarily; $g_{ij}(x)$ is the metric of the 3-spaces {t = const}. The second fundamental form is $\pi_{ij} = n_{i;j}$ where the normal to the surfaces {t = const} is $n_i = \delta^0_i$.

The matter flow lines have tangent vector $u^i = \delta^i_0$ (which differs from $n^i = g^{ij} n_j$ whenever $N^i \neq 0$). The shift vector $N^i(x^j)$ gives the change of the matter lines relative to the normal to the chosen time surfaces. The lapse function $N(x^i)$ gives the relation between coordinate time and proper time along the normal lines.

**The field equations** for $g_{ij}(x^k)$ are as follows (where 3-dimensional quantities have the prefix (3)): four constraint equations

$$^{(3)}R + \pi^2 - \pi_{ij}\pi^{ij} = 16\pi\ \rho_H \qquad (5)$$

$$R^\mu := -2\ \pi^{\mu j}{}_{|j} = 16\pi\ T^\mu{}_0 \qquad (6)$$

where "|j" represents the covariant derivative in the 3-surfaces, and twelve evolution equations

$$\partial_t g_{ij} = 2Ng^{-1/2}(\pi_{ij} - 1/2 g_{ij}\pi) + N_{i|j} + N_{j|I} \qquad (7)$$

$$\partial_t \pi_{ij} = -Ng^{-1/2}(^{(3)}R_{ij} - 1/2 g_{ij}{}^{(3)}R) + 1/2 Ng^{-1/2} g_{ij}(\pi_{mn}\pi^{mn} - 1/2\pi^2)$$
$$- 2Ng^{-1/2}(\pi_{im}\pi^m{}_j - 1/2\pi\pi_{ij}) + \sqrt{g}(N_{|ij} - g_{ij}N^{|m}{}_{|m}) + (\pi_{ij}N^m)_{|m}$$
$$- N_{i|m}\pi^m{}_j - N_{j|m}\pi^m{}_i + 16\pi \, ^{(3)}T_{ij}. \qquad (8)$$

What determines how the matter evolves? The matter energy-momentum conservation equations

$$T^{ab}{}_{;b} = 0 \qquad (9)$$

must be satisfied. Their outcome is determinate when equations of state for the matter terms $\rho_H$, $T^\mu{}_0$, and $^{(3)}T_{ij}$ in (5), (6), (8) are added, perhaps depending on internal variables such as temperature T, entropy S, or enthalpy H. These relations specify the physical properties of the matter. One may need additional dynamical equations for the matter to make the system determinate, as well as (9).

This can all be worked out using any time surfaces (that is the merit of the ADM formalism); in particular one can choose a unique gauge by specialising the time surfaces and flow lines to those defined above:

1. Choose the flow lines to be Ricci Eigenlines:

$$T^\mu{}_0 = 0 \Rightarrow R^\mu = -2\pi^{\mu j}{}_{|j} = 0 \qquad (10)$$

This algebraically determines the shift vector $N^i(x^j)$, so solving the constraint equations (6);

2. Determine the lapse function $N(x^i)$ by the condition that the time parameter t measures proper time $\tau$ along the fundamental flow lines:

$$ds^2 = -d\tau^2 \;\Rightarrow\; N^2 = 1 + N_i N^i \;. \qquad (11)$$

These conditions uniquely determine the lapse and shift. Then,

- given equations of state and dynamical equations for the matter, equations (7), (8), and (9) determine the time evolution of the metric in terms of proper time along the fundamental flow lines;
- the constraints (5), (6) are conserved because of energy-momentum conservation (9).

The development of spacetime with time thus takes place just as is the case for other physical fields, with the relevant time parameter being proper time $\tau$ along the fundamental flow lines. There is no problem with either the existence or the rate of flow of time. The spacetime develops accordingly via (7), (8).

**Predictability**: Do these equations mean the spacetime development is uniquely determined to the future and the past from initial data? That all depends on the equations of state of the matter: one can have an equation of state that involves random elements, as in the example in Figure 3.

The equations determine the time evolution of the spacetime, but do not guarantee predictability. Indeed if quantum unpredictability gets amplified to macro scales, the spacetime evolution is intrinsically undetermined till it happens as occurred for example during the generation of seed inhomogeneities in the inflationary era in the very early universe from quantum fluctuations[26-28], as originally pointed out by Mukhanov[35] (see Figure 4).

## 2.4 The issue of timelike surfaces

In an inhomogeneous realistic spacetime the surfaces S as determined above can become timelike in some regions, and then we need a timelike version of the initial value problem for that segment of S.

The same problem arises when initial data is set on a timelike Finite Infinity used to investigate isolated systems such as stars in general relativity[13]. Friedrich and Nagy[36] have developed the initial value problem for timelike surfaces in that context in the vacuum case, and the same methods should apply in this situation. Interesting geometric issues arise when S can possess both time-like and space-like subsets; however they should not be insuperable, as Ref.[36] shows.

## 2.5 The Issue of Scale

The above discussion in principle applies to all scales. However gravity is a very weak force, and only large masses have a significant effect on spacetime curvature. For that reason the scale to which the above discussion is relevant is Scale 3 in Table1: the scale of astronomical bodies is the scale that determines the structure of spacetime. Smaller scale entities have a negligible effect on this structure, rather entities at those scales just respond to spacetime curvature that is fixed at other scales. The spacetime geometry at cosmological size (scale 4) however is determined by coarse-graining or averaging[37] that at astronomical scales (Scale 3). This is discussed further in Section 4.

## 3. Quantum dynamics and non-unitary evolution

Quantum physics applies both to matter in the universe, and to the space-time structure of the universe itself. We consider them in turn.

### 3.1 The quantum dynamics of matter in the universe

Now we consider quantum dynamics in an existing spacetime structure, that is, Scale 1 dynamics (see Table 1). It is often claimed that quantum physics is unitary, hence the future is determinate. As stated above in Section 1.3, this ignores fundamental features of quantum theory. Quantum mechanics applied to the real universe does not only involve unitary transformations. Measurements happen; collapse of the wave function takes place; classical outcomes occur. This is not just decoherence, which effectively gets rid of entanglement but not superpositions.

**Measurement:** QM is experienced as non-unitary and irreversible when measurements take place, or more generally, wave function projection happens.[14,15,24,25] This is the core of the local flow of time: the indefinite future becomes the definite past as wave function collapse takes place. This happens all the time everywhere, it does not need to relate to an experiment. If a measurement of an observable $A$ takes place at time $t = t^*$, initially the wave function $\psi(x)$ is a linear combination of eigenfunctions $u_n(x)$ of the operator $\tilde{A}$ that represents $A$: for $t < t^*$, the wave function is

$$\psi_1(x) = \Sigma_n c_n u_n(x). \qquad (12)$$

But immediately after the measurement has taken place, the wave function is an eigenfunction of $\tilde{A}$:

$$\psi_2(x) = a_N u_N(x) \qquad (13)$$

for some specific value N. The data for $t < t^*$ do not determine the index N; they merely determine a probability $p_N$ for each possible outcome (13), labelled by N, through the fundamental equation

$$p_N = c_N^2 \qquad (14)$$

One can think of the projection of the initial state (12) to the eigenstate (13) as probabilistic time-irreversible collapse of the wave function, with probabilities given by (14). The initial state (12) does not uniquely determine the final state (13); and this is not due to lack of data, it is due to the foundational nature of quantum interactions- or at least it is indubitably the way QM is a determined to happen by experiments in a laboratory, whatever its foundations. You can predict the statistics of what is likely to happen but not the unique actual physical outcome, which unfolds in an unpredictable way as time progresses; you can only find out what this outcome is after it has happened.

This happens whenever a classical outcome occurs, for example a photo releases an electron in a rhodopsin or chlorophyll molecule; it does not depend on an observer. An example is the set of classical particle images appearing on the screen depicted in Figure 1 as individual particles arrive. Each such individual image is in effect a quantum measurement stating an electron arrived here and deposited energy at this particular place at a particular time. This completely resolves the initial uncertainty as to where the electron would arrive. Non unitary transformations certainly take place; ignoring this is ignoring a fundamental feature of physics.

We also can't retrodict to the past at the quantum level, because once the wave function has collapsed to an eigenstate we can't tell from its final state what it was before the measurement. Knowledge of the later state (14) does not suffice to determine the initial state (12) at times $t < t^*$, because the set of quantities $\psi_n$ are not determined by the single number $a_N$. Real quantum mechanics is not time reversible.

How does this relate to the EBU idea? It applies at Scale 1:

> ***Hypothesis: This is where the flow of time takes place at scale 1: the uncertainty of the future changes to the certainty of the past when collapse of the wave function (12) ➔ (13) takes place. This happens all the time everywhere***.

Why should the collapse of the wavefunction relate to the average motion of matter in the universe? - Because it is a contextually dependent effect, determined by the local physical environment, such as the measurement apparatus, or any physical system that causes collapse of the wave function (a screen, leaf, etc).[11] That higher level environment determines when the context for a wave function collapse has been set up. However this is a scale dependent statement: that average may look different at different scales (see Section 4).

**In summary:** Hamiltonian dynamics is not all that occurs in quantum physics. Irreversible non-unitary transformations (12) ➔ (13) also take place, and mark the change from probabilistic predictions to definite outcomes. If this did not happen, the wave function would have no meaning, as it would not predict anything via (14). No events would occur that would make up a statistical ensemble.

The events at Scale 1 then underlie emergence of structures and function at Scale 2, thus they underlie the emergence of time at that level.[11]

### 3.2 Quantum Gravity and Space Time Structure

As regards quantum gravity, the General Relativity results discussed in Section 2.3 (Scale 3) must emerge from an underlying quantum gravity theory (Scale 0). This presumably is due to collapse of the quantum gravity wave function to actualize the future from the present-boundary of the universe. When we have clarified in whatever way classical general relativity emerges from this as yet unknown theory of quantum gravity, that will help establish how the classical spacetime emerges from the underlying quantum gravity fields, whatever they are.

I will not speculate on these topics here except for making a few remarks.

**No universal wave function:** The following has been raised by a referee: "*if the universe as a whole (the present-front) undergoes collapse at the present boundary what does the collapsing, what is "measuring" the entire universe? In other words, which interpretation of quantum presupposed here?*"

This paper assumes the standard view of collapse of the wave function as discussed in Section 3.1 applies even to quantum gravity, with the following nuance: nothing is measuring the entire universe, indeed there is no useful universal wave function. Rather local wave functions (at scales 0 and 1) exist everywhere describing local systems, and local measurements take place based on these local wave functions; larger scale effects (at scales 2 and 3) emerge from these smaller scale effects by coarse-graining[11]. Insofar as a global wave function exists, it is an emergent state arising out of all the local wave functions that exist, and its dynamics derive from that feature. It will not be expected to evolve in a unitary way because non-unitary local measurements of local wave functions take place, and that will generically lead to a non-unitary evolution of their tensor product. No "measurements" of this wave function as a whole takes place: there is no context in which that could occur. It evolves through bottom-up processes.

Those who disagree with this proposal are invited to propose an experiment that would prove existence of a wave function of the universe which evolves in a unitary way (the proposal that it does exist is an extraordinary extrapolation from laboratory scale to the scale of the whole universe: such an extrapolation surely needs experimental exploration, else it is just a philosophical speculation).

**A preferred reference frame:** "*If collapse is supposed to provide a preferred frame, that needs to be justified, there is nothing inherent about collapse that provides a preferred frame.*" This puts the thing the wrong way round. It is the local context that defines local preferred frames: they then set the context for wave function collapse to occur, as discussed in Ref. [11]

**A consistency condition:** The proposal here will be that whatever foundations are set at the quantum gravity level, they must lead to emergence of an EBU at the macro level, or it is an inadequate theory. This is to be taken as a selection rule for theories of quantum gravity.

Indeed quantum gravity can be based in EBU-like models, such as spin foam models based in a discrete spacetime picture,[38,39] which would seem a promising approach.

4. **The Meshing of Scales**

The proposal made here is to use proper time along preferred timeline as the time of evolution for the EBU, but the issue then is on what scale of description? The answers have been implicit above. The point that emerges is we must distinguish between emergence of the spacetime itself, and the concretization of events within spacetime.

- As far as spacetime is concerned, the effective time fixing the EBU structure must be determined on the scale that controls space time structure, that is, this determination must take place on the basis of the matter distribution at Scale 3, even though the wave function collapse leading to its existence must be based in quantum gravity processes (Scale 0).

- As far as macro entities at Scale 2 are concerned, they experience the proper time determined by the space time structure, but have negligible influence on that structure. The present time for them, when the indeterminate future concretizes to the determinate past at this scale, must obviously lie within the space-time domain that has come into existence, but need not necessarily coincide with the gravitational present. It will emerge from the underlying quantum (level 1) structures, where, because of evidence provided by delayed choice experiments[45-47], the emergent present may be expected to have a crinkly nature characterized as a Crystalizing Block Universe (CBU) structure (this is discussed by Ellis and Rothman[2]).

The proper time for individual observers who move relative to the universal rest frame defined at scale 3 will not necessarily coincide with the universal proper time. But this is just the same as in special relativity: individual proper times between spacetime events, determined at scale 2, differ according to the path traveled between them. Time dilation takes place between universal time (scale 3) and local times (scale 2). It is the former that determines the spacetime structure. However the determination of the effective present at scale 2 is based in quantum processes at level 1.

Accordingly the implication is we must distinguish *Global Time* – related to emergence of spacetime itself and determined by Level 3 entities, from *Local Time* – which relates to the experience of time at the macro level (Level 2), and is emergent from the underlying Level 1 entities. These two are logically distinct from each other, because they have different physical bases. It is possible in GR to have twins part company and rejoin to shake hands at different ages. Their ages will in general not coincide with the macro present, which is determined by an average motion of matter at the macro scale 2, because this sets the context for wave function collapse events at scale 1 to take place (see Ellis[11]).

**Consistency conditions.** These need careful consideration: here are a few preliminary remarks.

- The basic consistency requirement is that the local time for moving observers cannot give greater times than global time. The surface where local time crystalizes must lie within the evolving future boundary of the EBU.

- Level 2 time emerges from level 1, where time is crinkly, as pointed out by Ellis and Rothman[2]: some local regions crystalize out later than others, due to delayed choice effects. How this happens can only be clarified when we have a good theory of wave function collapse.

The main point I make here is that as remarked above, we can expect wave function collapse to be in some sense controlled by the local Scale 2 context. [11] This should lead to consistency between Level 1 and Level 2 views of the present. However Level 3 time emerges from Level 0, but no one has a theory of wave function collapse at the quantum gravity level: the issue is not even discussed in quantum gravity texts, except perhaps to propose the Everett many worlds view. But that view is incompatible with the usual Block Universe picture, as pointed out in Section 1.2.

A very interesting topic is how this all relates to the psychological perception of time, where many experiments show various apparent anomalies. That is a great topic for future investigation.

5. **The direction of time and the arrow of time**

Microphysics interactions (except for some weak interactions which have a negligible effect on daily life) are time symmetric: the future and past are equal as far as electromagnetism and gravity are concerned. How can a difference emerge at the macro scale between the future and the past, on the basis of time symmetric microphysics? How does time know which way to flow? Why does it flow the same way everywhere?[17,43-45] In the block universe the two directions of time are equal. Not so at scale 2, according to the second law of thermodynamics!

There is no basis for a determination of the arrow of time in unitary microphysics alone: Boltzmann's proof of the H-theorem **dS/dt > 0** by coarse graining microphysics applies equally in both time directions (set t ➔ t' = -t: exactly the same derivation of the H-theorem will hold and show **dS/dt' > 0** also[17]). The same applies to the Quantum Field Theory derivation of the Second Law by Weinberg[46]: it predicts **dS/dt > 0** in both directions of time. Neither derivation provides a foundation for the second law of thermodynamics with a unique arrow of time.

In what follows I distinguish between the global direction of time, and local arrows of time.[4]

**The EBU sets a direction of time** The evolving block universe provides a cause of a unique direction of time: namely the past exists, and is developing to the future, which does not yet exist.

- The future can't affect us today, because it is not yet definite what it will be. Possibilities exist, constrained by conservation laws, but not specific outcomes. At the macro scale, causal effects do not reach back to us from the future.

- The past affects us today in many ways: for example the heavy elements on Earth, cosmic ray particles causing genetic mutations, and the cosmic background radiation we detect at the present time all originated in the determinate past.

The direction of time arises fundamentally because the future does not yet exist, but the past does: a global asymmetry in the physics context. One can be influenced at the present time from many causes lying in our past, as they have already taken place and their influence can thereafter be felt. One cannot be physically influenced by causes coming from the future, for they have not yet come into being. This is the rationale for saying the past exists but the future does not: if something can influence you, it exists. The direction of time is non-locally determined: it points from the start of universe, which a fixed unchanging past boundary to spacetime, to the ever-changing present, where the future endpoint of spacetime is continually extending to the future.[4]

**Special initial conditions sets the thermodynamic arrow of time** The electromagnetic and quantum field theory arrows of times must necessarily be aligned with the global direction of time provided by the EBU, because only retarded Green's functions can have meaning in the EBU context. We must however have had special initial conditions at the start of the universe (a `past condition') in order that the second law of thermodynamics holds in the forward direction of time.[17,44,45] This then aligns the other local arrows of time (chemistry, biological, and the mind) with the direction of time provided by the evolution of the universe. The arrows of time cascade down to lower levels of scale by top down effects, and then up through complex structures by emergent effects[4].

6. **Chronology Protection**

A longstanding problem is that, as demonstrated by Gödel, closed timelike lines can occur in exact solutions of the Einstein Field Equations with reasonable matter content.[31] This opens up the possibility of many paradoxes, such as killing your own grandparents before you were born and so creating causally untenable situations. It has been hypothesized that a Chronology Protection Condition would prevent this happening.[47] This is however an add on to the Einstein Field Equations: an ad hoc condition added on as an extra requirement on solutions of the Einstein field equations, which do not by themselves give the needed protection.

The EBU automatically provides such protection, because creating closed timelike lines requires the undetermined part of spacetime intruding on regions that have already been fixed. This would require the fundamental world lines defined at scale 3 to intersect; but if these fundamental world lines intersect, the density diverges and a spacetime singularity occurs. The worldlines are then incomplete and time comes to an end there; so no "Grandfather Paradox" can occur because there are no relevant world lines that individuals can move on. Hence the EBU as outlined above automatically provides chronology protection. Indeed we get only spacetimes with global time orientation in the Cauchy development of the initial data surface.

Two comments are in order here.

Firstly, individual macro objects cannot cause spacetime curvature leading to closed timelike lines, as in the case of the Gödel universe[31], because they are not large enough to significantly curve spacetime. They can only explore the spacetime that exists; and in an EBU that cannot have closed timelike lines, as just explained. Hence there is no danger of a meeting of individual persons causing a singularity.

Secondly, a paper[48] considers quantum theory experiments to test for existence of Closed timelike curves (CTCs), stated to be trajectories in spacetime that effectively travel backwards in time. However this paper does not relate to CTCS created by the global spacetime structure: rather it relates to trajectories in Minkowski spacetime, where as already remarked, quantum field theory allows effects that appear as if they relate to travel into the past[40-42]. This relates to the idea of existence of a Crystallizing Block Universe[4] at scale 1, which is compatible with what is discussed here.

### 7. A more realistic view, and objections

Physics needs to take everyday reality into account; hence whatever the micro foundations may be, they must lead to emergence of an EBU at the macro level (scale 2). It must also take quantum physics seriously, so there must be an EBU at the micro level (scale 1), see Figure 1. This then leads to an EBU at the astronomical level (Scale 3) because of the inflationary phase in the early universe, see Figure 2.

### 7.1 Objections

The main objection to the EBU (the lack of preferred surfaces of change) has been answered above. A further set of objections have been made to the proposal, which I now answer.

**Objection 1: Time does not flow**. It has been claimed that "Time does not flow: this is incoherent." This is correct: time does not flow, which suggests it is moving past something else, it passes, meaning that the future boundary of space time ever includes more events on an ongoing basis as the potential of the future becomes the reality of the past. It is the passage of time that allows rivers to flow and other events to take place.[49]

**Objection 2: Can't have a rate**. A key question is, "How fast does time pass?" Davies[9], and others suggest there is no sensible answer to this question. I claim that the answer is given by the metric tensor $g_{ij}(x^k)$, which determines proper time $\tau$ along any world line by equation (2). This is the time measured locally along that world line by any perfect clock. This is what fixes physical time, including gravitational time dilation [gravitational potential affects relative rates], along world lines. Real world clocks - oscillators that obey the Simple Harmonic Equation - are approximations to such ideal clocks, and it is the relation between such clocks and other physical events that measures the passage of time. The global rate derives from the combination of these local rates.

A more specific version of this claim is the statement "Time can't pass at the rate of one second per second because that's not a rate it's a dimensionless number". This is wrong. The situation is just like rates of exchange of money: this is an operator with two slots, each with its own units, they don't cancel, as pointed out by Maudlin[49]. Hence time flows at the rate of one second per second, as determined by the metric tensor locally at each event. There is no inconsistency.

**Objection 3. Not necessary to describe events** Davies[9] and Rovelli[7] claim time does not pass because it's not needed to describe the relations between relevant variables, which are all that matter physically. Thus you can always get correlations between position p(t) and momentum q(t) for a system by eliminating the time variable: solve for t = t(q) and then substitute to get p(t) = p(t(q)) = p(q), and time has vanished! Thus time may exist but it does not flow; only correlations matter.

But the latter model leaves out part of what is happening: that does not mean it does not happen, it just means it's a partial model of reality, including some aspects and omitting others. It's a projection from spacetime to a phase portrait. It leaves out the way that the continually changing correlations flow smoothly one after another in a continuous ongoing way. Take for example Kepler's three laws of planetary motion. If only relations between dynamic variables count, not time, one reduces to Kepler's one law and miss two thirds of his discoveries! Yes of course it's relative to clocks. Their ticking measures the passage of time. Without the passage of time, they don't tick

**Objection 4: Categorization problem** A philosophical argument by MacTaggart[50] and Price[1] is that the past, present, and future are exclusive categories, so a single event can't have the character of belonging to all three.

The counter is as follows: Suppose E happens at $t_E$.
- At time $t_1 < t_E$, E is in future,
- At time $t_1 = t_E$, E is in present,
- At time $t_1 > t_E$, E is in past.

Its category changes - that is the essence of the flow of time - so this is a semantic problem, not a logical one. One needs adequate semantic usage and philosophical categories to allow description of this change: language usage can't prevent the flow of time!

**Objection 5: Unitary dynamics** The claim is made that spacetime develops according to Hamiltonian (unitary) dynamics[6]. I have dealt with this above – this is only true if you ignore collapse of the wave function at the micro level, and all irreversible processes at the macro level, in essence claiming they are not really irreversible. Experience dictates otherwise. That is not true in terms of descriptions at each of those levels.

A particular case is the claim that the Wheeler de Witt equation, which is essentially the Hamiltonian constraint turned into a quantum operator acting on `the wave function of the universe', only gives time independent solutions, hence time does not flow. However this proposal is problematic (see[51]):
- This argument ignores the issue of measurement, without which the wavefunction has no meaning unless one tries to go the Everett multiverse path, which has great difficulty with realising the Born rule (14) as well as a preferred basis problem;
- The equation has definitional and divergence problems;
- There is no experimental evidence that the equation actually applies to any real physical systems in any context whatever;
- There is no evidence that quantum theory applies to the entire universe, as is implied by the idea of the wave function of the universe. This is an extraordinary extrapolation from the microphysical domain where quantum theory has been shown to apply.

In the end the fact that time does indeed pass shows that this equation does not by itself adequately represent the dynamics of space time structure.

**Objection 6: Delayed Choice Experiments**[40-42] indicate that quantum effects can to some degree reach back into the past, and the EBU does not take this into account. The response is that the EBU picture can be modified to the Crystalizing Block Universe idea, where local delays in actualisation take this into account, see Ref.[2]

**7.2 A more realistic view**

Physics needs to take everyday reality into account. Did it make sense for the Planck team to announce the present age of the universe as 13.784 Gyr on March 21, 2013?[52] And is the universe at the time of writing a bit more than one year older?

This article proposes that both of these make sense, because spacetime is an evolving block universe, with the present the future boundary of spacetime which steadily extends in to the future as time progresses. The present separates the past (which already exists) from the future (which does not yet exist, and is indeterminate because of foundational quantum uncertainty). To consider this properly, one must carefully consider the various scales that characterise physical structure, and distinguish global and local time, as discussed above.

There are some technical aspects to this - namely
   (1) Simultaneity is a purely psychological construct,
   (2) One can define unique surfaces of change in a non-local geometric way,
   (3) This structure prevents existence of closed timelike lines,
   (4) The arrow of time is distinguished from the direction of time, which is non-locally defined in this context.

There are issues needing development:
   (5) The first technical issue that needs development is how the initial value problem works when these surfaces become timelike. This is an unusual situation that will need careful thought, but initial work by Friedrich and Nagy[36] shows this can be done.
   (6) The second technical issue needing development is careful further consideration of how this works out at the various levels of structure and scale, and how they relate to the passage of time (as above) and the arrow of time (as in Ref[4]).
   (7) The final major issue needing development is proposing a viable theory of wave function collapse. Various initial proposals are promising in this regard, but need improvement.

**Acknowledgement:**

I thank an anonymous referee for challenging comments that have led to a great improvement in this paper.

-----------------------------------------------------

Figure 1

Double-slit-experiment performed by Dr. Tonomura, showing the build-up of an interference pattern of single electrons. The numbers of electrons are (a) 200, (b) 6000, (c) 40000, (d) 140000 (Wikimedia Commons). The individual events occur at intrinsically random places: there is no cause for where they appear on the screen. However the statistical pattern that thereby is built up is completely determined

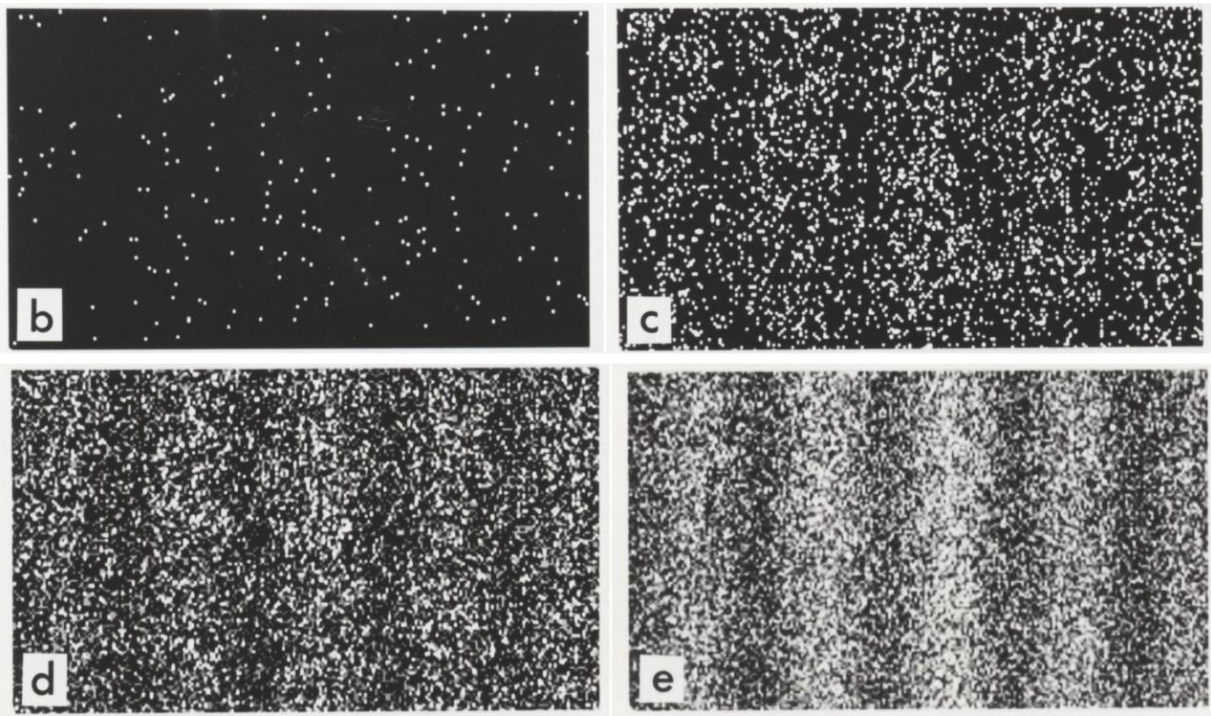

**Figure 2**

According to the standard inflationary picture, the present day large scale structure in the universe is the outcome of quantum fluctuations during inflation. Hence the specific individual galaxies that occur are not determined by initial data at the start of inflation

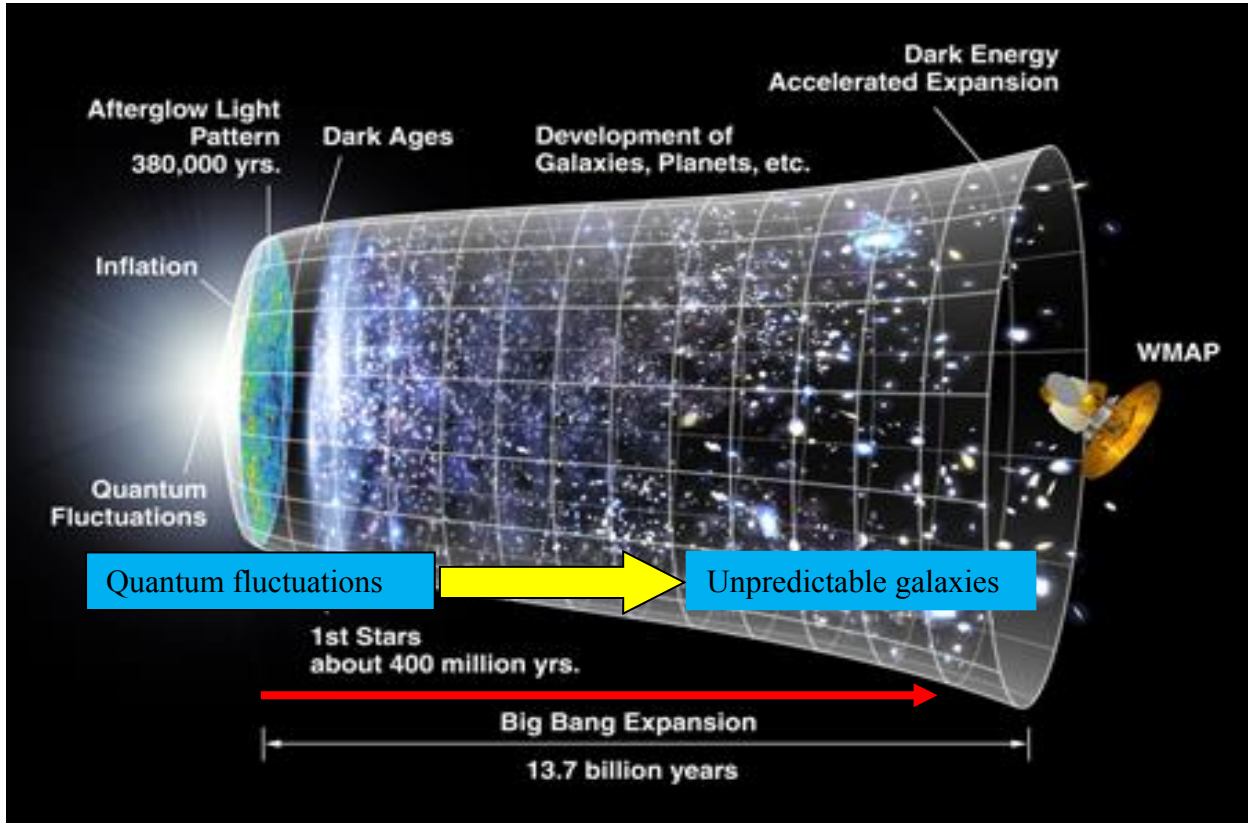

**Figure 3**

Space-time diagram of massive object driven by computer controlled rocket motors in an unpredictable way. On the left, the situation at time $t_1$: there are many paths possible. On the right, the situation at time $t_2$: all but one option at time $t_1$ have been rejected, one path has been chosen. A further set of options are open at time $t_2$.

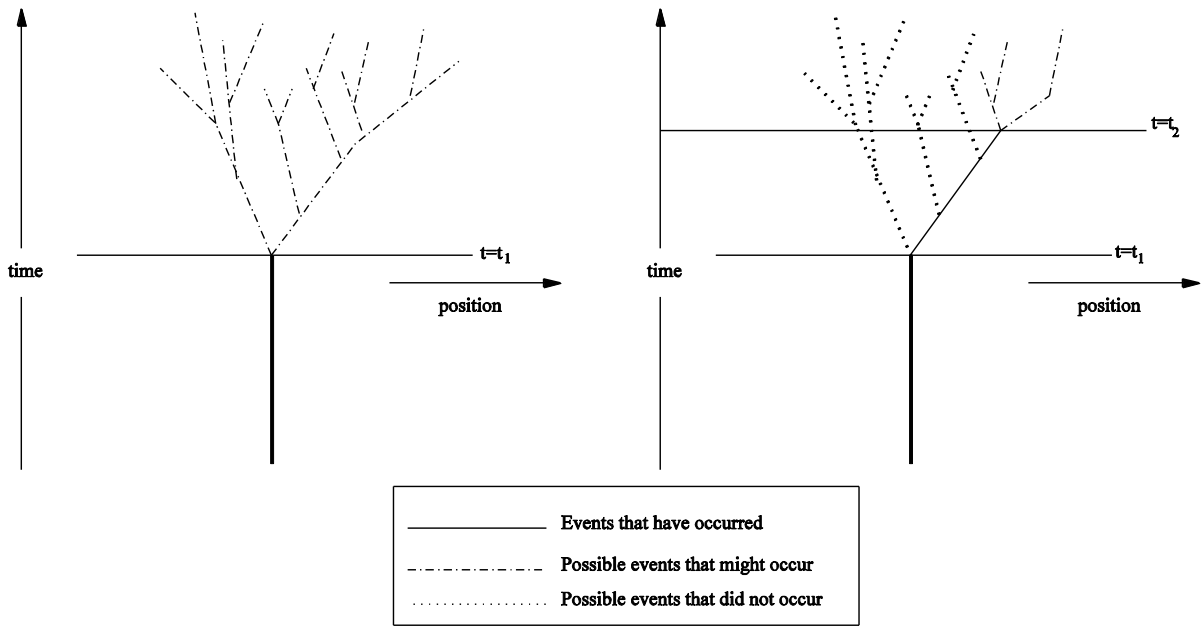

**Figure 4**

The evolving block universe: a spacetime that is at each instant bounded to the future by the ever changing present time. As time passes, the future boundary of spacetime extends to include more events; the initial boundary (the start of the universe) is fixed and unchanging. This extension takes place along preferred timelike world lines.

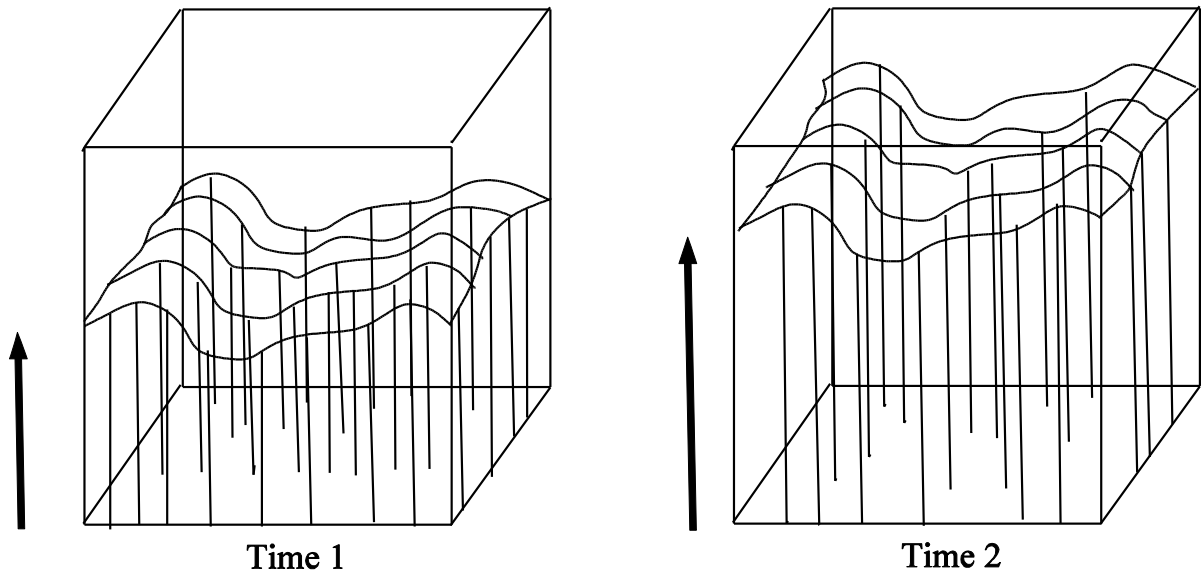